\documentclass[twocolumn,showpacs,preprintnumbers,amsmath,amssymb,superscriptaddress,groupedaddress,prb]{revtex4-1}

\usepackage{epsfig}                                                            
\usepackage{graphicx}
\usepackage{dcolumn}
\usepackage{color}
\usepackage{bm}
\usepackage{subfigure} 
\usepackage{epsfig} 

\begin{document}

\title {\bf Current-driven orbital order-disorder transition in LaMnO$_3$  }




\author{Parthasarathi Mondal}
\email{Currently at the Department of Advanced Material Science and Engineering, Sungkyunkwan University, Suwan 440-706, Korea} \affiliation{Nanostructured Materials Division, Central Glass and Ceramic Research Institute, CSIR, Kolkata 700032, India}
\author{Dipten Bhattacharya} \email{Corresponding author; dipten@cgcri.res.in}\affiliation{Nanostructured Materials Division, Central Glass and Ceramic Research Institute, CSIR, Kolkata 700032, India}
\author{P. Mandal} \affiliation{Experimental Condensed Matter Physics, Saha Institute of Nuclear Physics, Kolkata 700064, India}

\date{\today}

\begin{abstract}
We report a significant influence of electric current on the orbital order-disorder transition in LaMnO$_3$. The transition temperature \textit{T}$_{OO}$, thermal hysteresis in the resistivity ($\rho$) versus temperature(\textit{T}) plot around \textit{T}$_{OO}$, and latent heat \textit{L} associated with the transition decrease with the increase in current density. Eventually, at a critical current density, \textit{L} reaches zero. The transition zone, on the other hand, broadens with the increase in current density. The states at ordered, disordered, and transition zone are all found to be stable within the time window from $\sim$10$^{-3}$ to $\sim$10$^4$s.  
\end{abstract}        

\pacs{71.70.Ej, 64.60.Cn, 71.30.+h, 71.27.+a}
\maketitle    
\section{Introduction}
The long-range orbital order in LaMnO$_3$ develops with ordering of active Mn $3d_{3x^2-r^2}$ and $3d_{3y^2-r^2}$ orbitals, alternately at each Mn site, within a Mn-O plane and the stacking of this order along c-axis (d-type order).\cite{Mizokawa} It undergoes a reversible order-disorder transition at a characteristic transition temperature T$_{OO}$.\cite{Murakami} The orbital order superstructure originates from cooperative fluctuations of the doubly degenerate Mn 3d$e^1_g$ orbitals interacting via Kugel-Khomskii superexchange. This is further aided by cooperative Jahn-Teller distortion of the Mn$^{3+}O_6$ octahedra.\cite{Kugel,Okamoto,Pavarini} The structurally forbidden orbital Bragg peaks could be clearly observed, with expected azimuthal angle dependence of peak intensity, in resonant x-ray scattering experiment, thus offering decisive evidence for formation of orbital order superstructure in LaMnO$_3$.\cite{Murakami} The orbital order, of course, is not a continuum but contains domains due to interaction with intrinsic lattice strain and/or defects.\cite{Ahn} This orbital domain structure could also be observed in spatially resolved coherent x-ray scattering experiment.\cite{Nelson} It is both technologically as well as fundamentally important to explore whether or not such an orbital ordered structure undergoes an order-disorder transition upon electric, magnetic or optical stimulations. It has been shown in the past that the long-range charge order in the doped systems melts down under electric, magnetic, and optical stimulations yielding sharp rise in magnetization and/or conductivity together with change in the crystallographic structures.\cite{Tomioka,Asamitsu,Miyano,Kiryukhin,Yamanouchi,Hervieu,Rao} The orbital stripes too were shown to undergo rotation under electric field in charge/orbital ordered layered manganites.\cite{Hatsuda,Polli} While in LaMnO$_3$, the Jahn-Teller distortion and orbital order was found to quench completely under $\sim$20 GPa mechanical pressure,\cite{Loa} in LaVO$_3$, the high energy laser pulse could melt the orbital order.\cite{Tomimoto} In spite of such rich background, there is, as yet, no information about whether or not in $\textit{undoped}$ LaMnO$_3$ the orbital order-disorder transition can be driven by an electric field or photoirradiation. More importantly, in none of the past experiments on external-stimulation-driven phase transition, has an attempt been made to track the evolution of the order of transition, i.e., how the latent heat of the transition, if any, varies with the increase in electrical, magnetic, optical, and mechanical energy, directly by calorimetry.

In this paper, we report the observation of significant influence of an electric current on orbital order-disorder transition in LaMnO$_3$. Using the experimental data, a phase diagram on the current density (\textit{J})-temperature (\textit{T}) plane is constructed. We find that the latent heat (\textit{L}) associated with the transition becomes zero at a critical current density $J_C$ $\sim$ 50 A/cm$^2$. It is also observed that the transition width broadens continuously as \textit{J} is increased. The zones in the \textit{J-T} plane - at well below transition, within the transition region, and at well above the transition - are characterized by probing their resistivity relaxation behaviors together with the resistivity (\textit{$\rho$}) versus temperature (\textit{T}) patterns. 
\section{Experiments}
The experiments are carried out on high quality single crystals of LaMnO$_3$ of dimensions 5 $\times$ 3 $\times$ (0.5-1.5) mm$^3$.\cite{Mandal} The gold electrodes and wires are used in standard linear four-probe configuration for the measurements of $\rho-T$ and $\rho$ versus time (at a given temperature) patterns under different bias currents. During the measurement of resistivity, the temperature sensor is attached directly onto the surface of the sample in order to record the data as a function of actual sample temperature. We also record the differential thermal analysis (DTA) data simultaneously with $\rho-T$ measurements under varying bias current in order to estimate the latent heat of the transition, from the DTA peak around the transition point, as a function of applied current density. We report all the data as a function of the applied current density as the measurements are done by passing different bias current directly through the sample.

\begin{figure*}[htp]
  \begin{center}
    \subfigure[]{\includegraphics[scale=0.45]{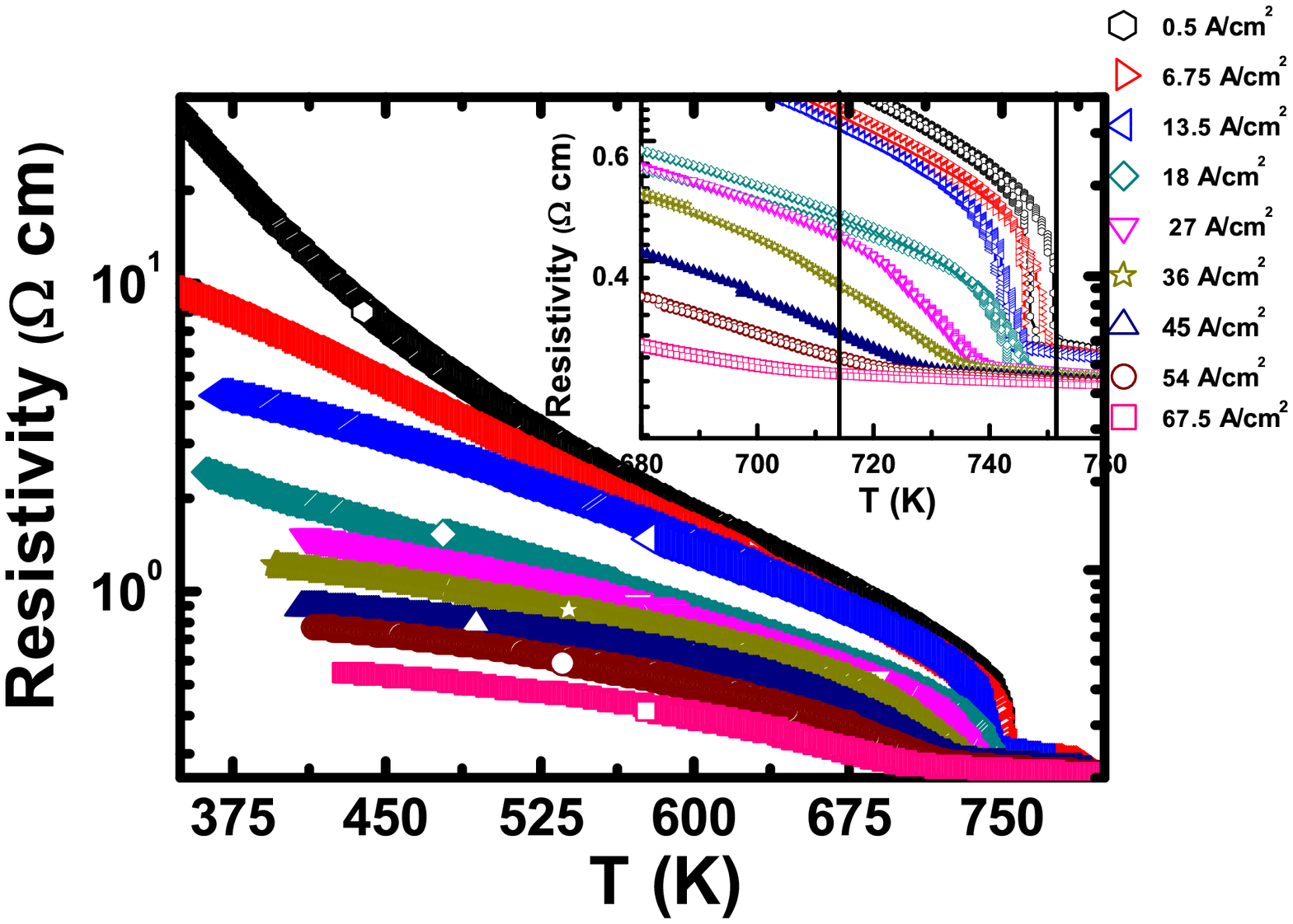}} 
    \subfigure[]{\includegraphics[scale=0.40]{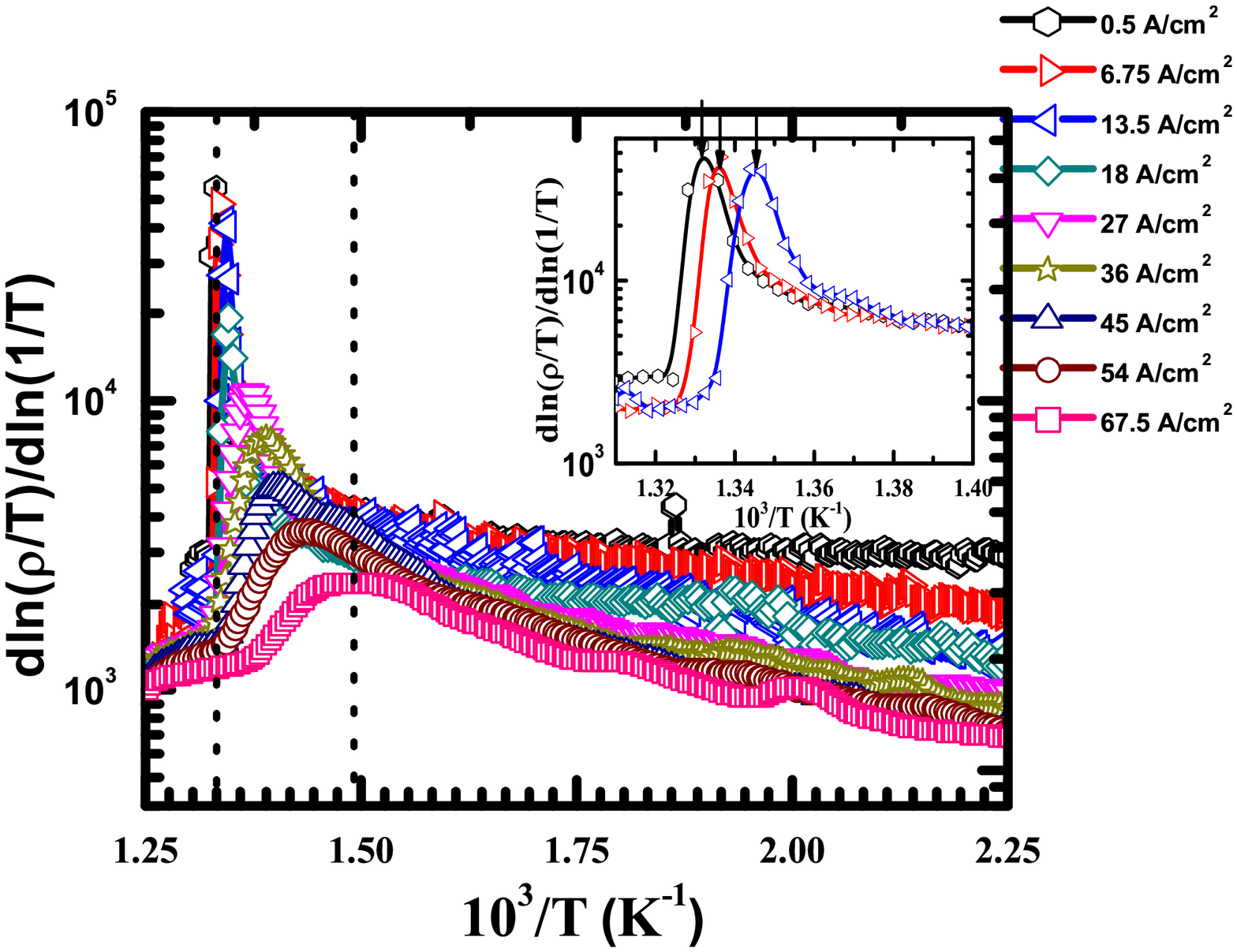}}
    \subfigure[]{\includegraphics[scale=0.42]{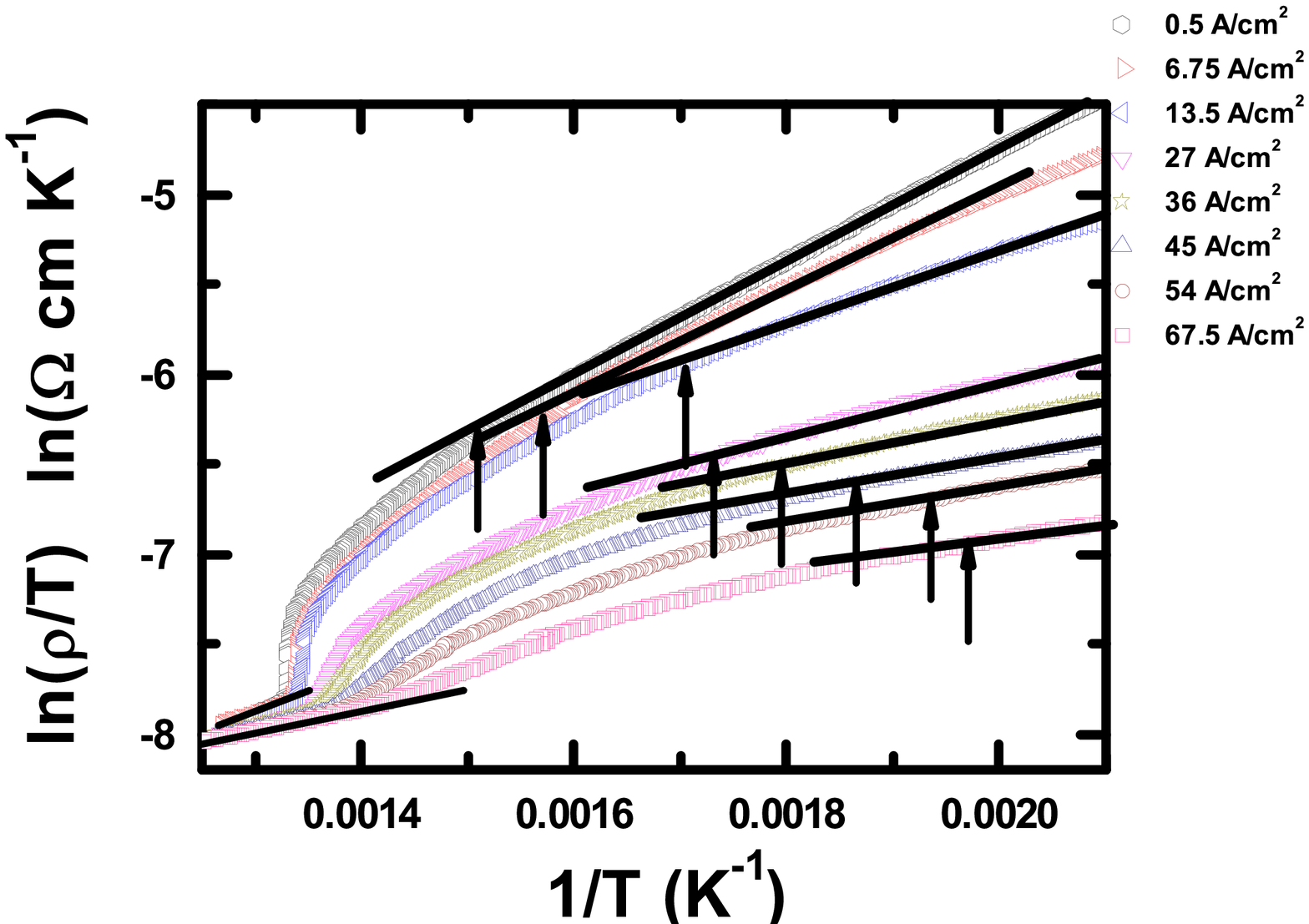}}
    \subfigure[]{\includegraphics[scale=0.40]{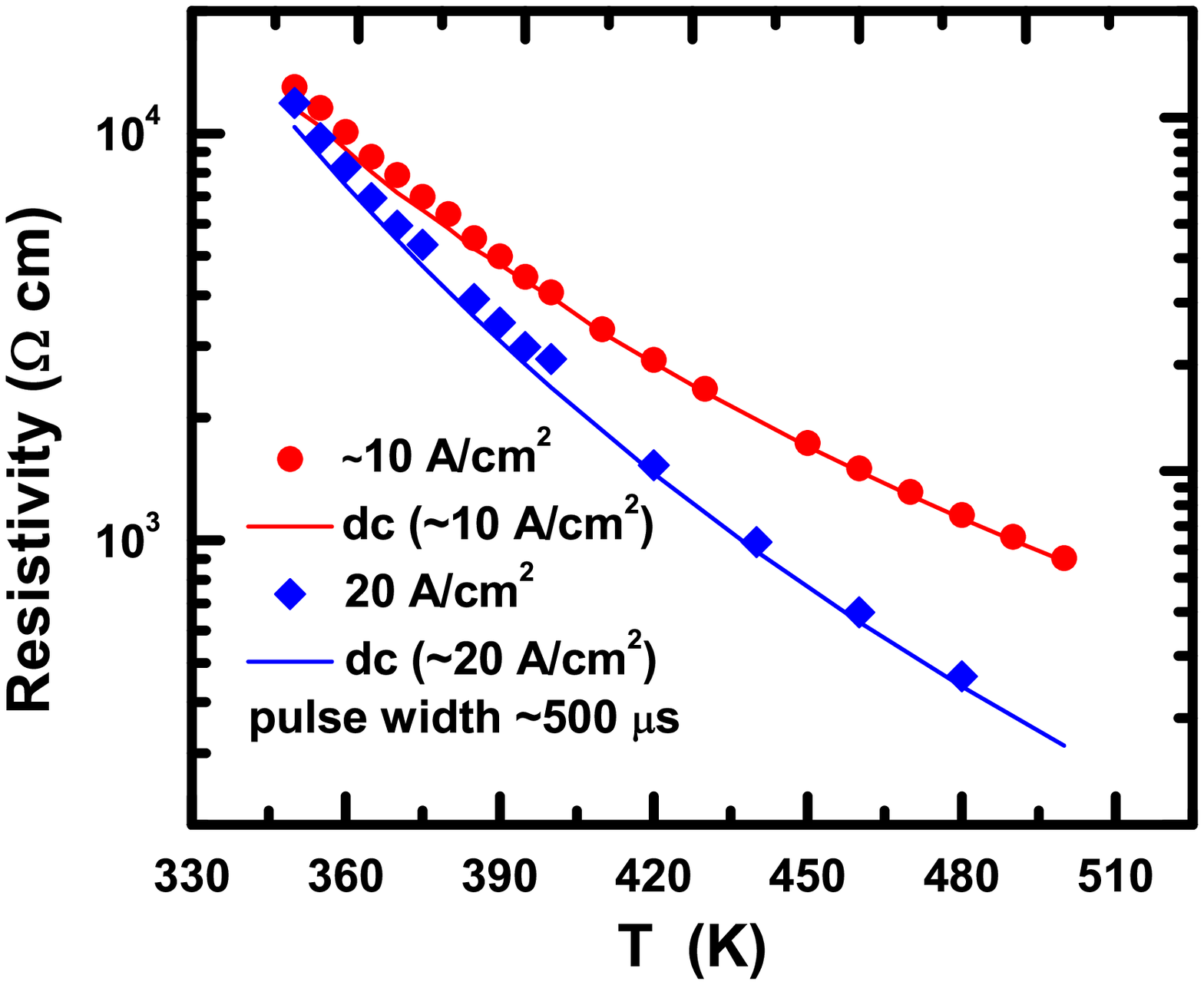}}  
  \end{center}
  \caption{(color online) (a) The resistivity ($\rho$) versus temperature (T) plot under different bias current at above room temperature showing the variation in T$_{OO}$ with the bias current density; inset: the region around T$_{OO}$ is blown up; (b) The $dln(\rho/T)/d(1/T)$ versus $1/T$ plots for different bias current density obtained from the resistivity ($\rho$) versus temperature (T) data; the temperature corresponding to the peak is the T$_{OO}$; inset shows how the T$_{OO}$ shifts with the increase in current density in low current density regime; (c) $ln(\rho/T)$ versus $1/T$ plots; the onset of the transition (T$^*$) is marked by arrow; (d) comparison of the data obtained from continuous dc and pulsed current measurement is shown for two representative cases. In all the plots the current density increases as one moves from top to bottom. }
\end{figure*}  

\section{Results and Discussion}
In Fig. 1(a) we show the $\rho$ -$T$ data measured under different bias current. It is quite evident from the inset of Fig.1(a) that the transition temperature \textit{T}$_{OO}$ decreases under increasing current (or bias voltage). In Fig. 1(b) we plot the $d\ln(R/T)/d(1/T)$ versus $1/T$ patterns. The peak in this plot marks the \textit{T}$_{OO}$; the height of the peak decreases while the width increases as the bias current increases. Figure 1(c) shows the deviation of the R-T pattern from adiabatic small polaron hopping model $R = R_0T^{\alpha}.\exp(\Delta /k_BT)$ ($\alpha = 1$) beyond a certain temperature \textit{T$^*$} (marked in the figure by arrow) below \textit{T}$_{OO}$. T$^*$, therefore, marks the onset of the transition. An error of $\pm$0.5\% is, of course, estimated to be involved in identifying the \textit{T}$^*$. Like \textit{T}$_{OO}$, \textit{T}$^*$ also decreases progressively with the increase in bias current density. The zone confined within \textit{T}$^*$ and \textit{T}$_{OO}$ marks the transition zone where both orbital ordered and disordered phases are expected to coexist and the transport of charge carriers does not follow any model applicable to motion with long range coherence.
An important issue here is the Joule heating of the samples which renders the determination of the current driven intrinsic effects difficult. In order to quantify the impact of Joule heating under enhanced bias current (or electric field), we measure the rise in temperature due to heating by attaching a temperature sensor directly onto the sample surface. Such an arrangement has earlier been used by others for measuring the actual sample temperature governed both by heating from the bath and Joule heating.\cite{Carneiro} We found, as expected, that the rise in temperature, as a result of current flow through the sample and sample-current lead junctions, enhances with the enhancement of current density J: from nearly negligible to as high as 100 K for a current range 50 $\mu$A - 1A. We start the current flow through the sample at room temperature and allow the temperature to rise by a certain extent within that range. Once the temperature stabilizes at a particular point above room temperature, we start the measurement. At that point no difference between the furnace (i.e., bath) and actual sample temperature could be noticed. The difference reduces to zero as the sample resistance and, therefore, the Joule heating drops down drastically. The actual sample temperature has all along been monitored while recording the $\rho -T$ data. Similar dc current driven measurements on charge ordered compound has earlier been carried out by others and it was found that the impact of Joule heating is minimum.\cite{Rao} In order to quantify the influence of Joule heating even further, we have compared the dc resistivity pattern around T$_{OO}$ with the pattern obtained under pulsed current (pulse width $\sim$500 $\mu$s). Both the data set are found to be nearly identical in the high temperature regime (Fig. 1d) establishing negligible role of Joule heating in that zone. Moreover, reproducibility of the features of transition, such as, broadening of the peak in the DTA pattern, decrease in the hysteresis and jump in $\rho -T$ around the transition, close matching of data between heating and cooling cycles etc in crystals of different thickness reveals that these are intrinsic field-driven effects. The experiments have been repeated on crystals of different thickness and all the features of transition were found to be reproducible. The Joule heating cannot give rise to these features, reproducibly, around T$_{OO}$. 

\begin{figure}[!hbp]
\centering
\includegraphics[scale=.3]{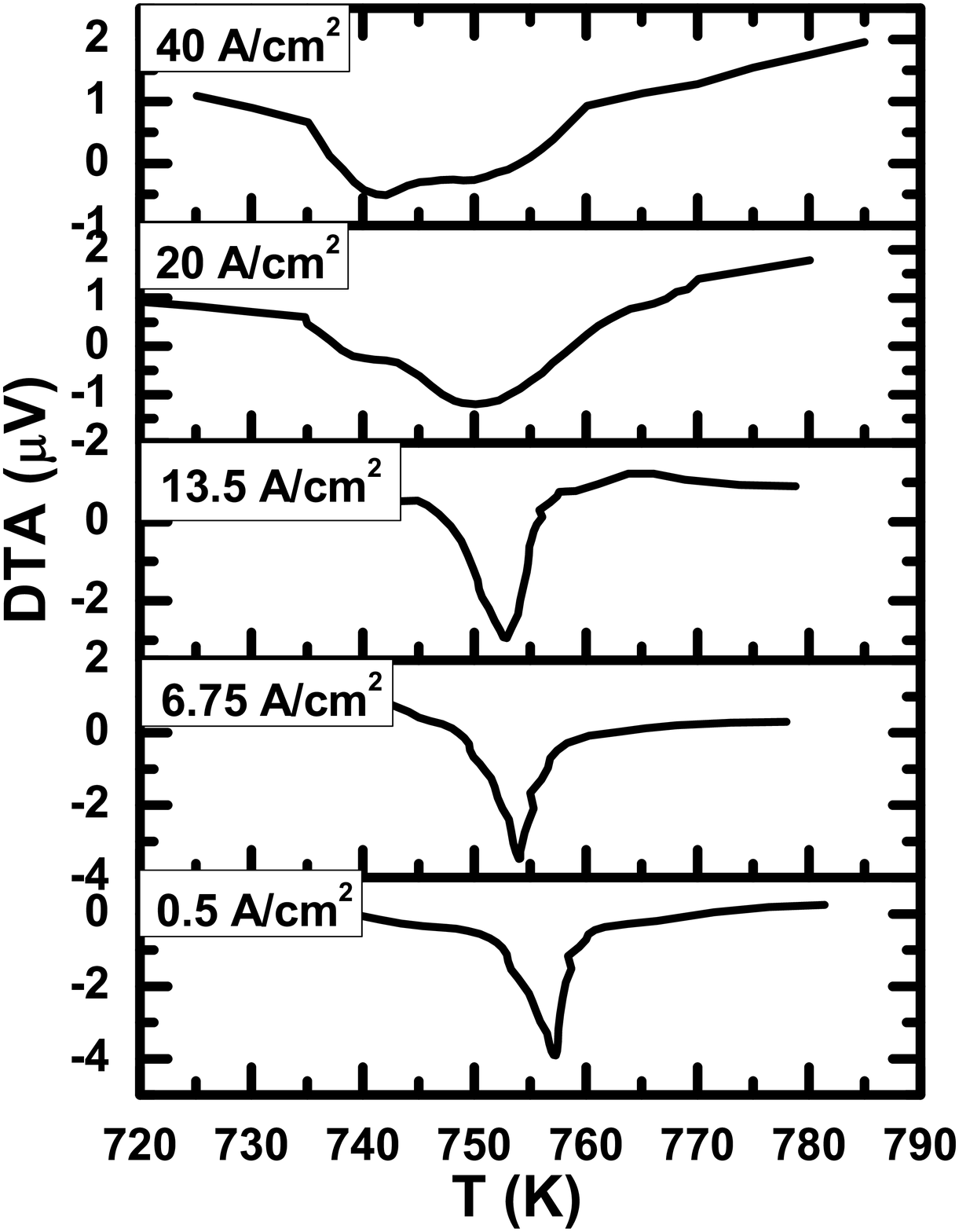}
\caption {The representative raw DTA traces in heating cycle, observed under different bias current, showing endothermic peaks around the orbital order-disorder transition; the peak shifts and broadens while the area decreases as the bias current increases. } 
\end{figure}

\begin{figure}[!hbp]
  \begin{center}
    \subfigure[]{\includegraphics[scale=0.43]{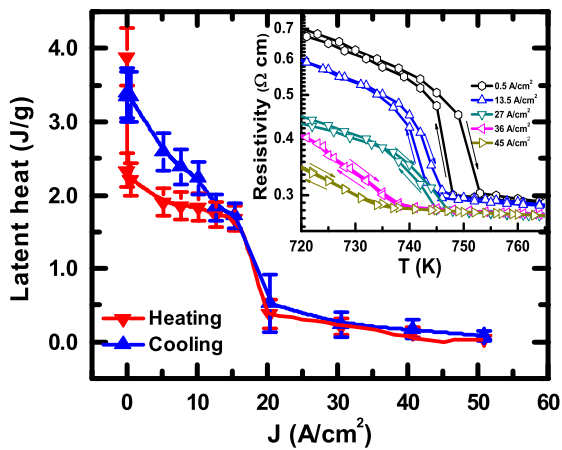}} 
    \subfigure[]{\includegraphics[scale=0.40]{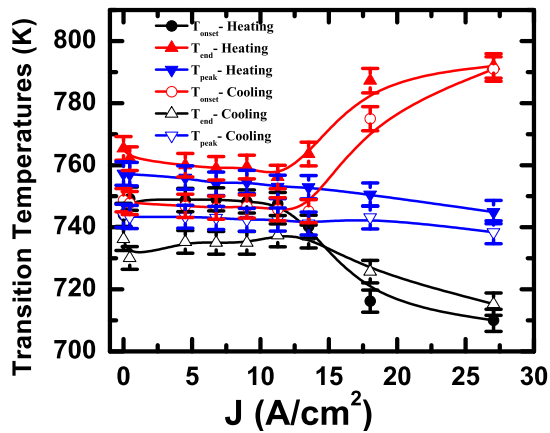}}
  \end{center}
  \caption{(color online) (a) The variation of latent heat-estimated from the area under the peak observed in DTA thermogram near the orbital order-disorder transition point-with bias current density (J); inset shows how the extent of hysteresis decreases with the increase in bias current density; current density increases as one moves from top to bottom of the inset plot; (b) variation of the transition temperatures (noted from DTA thermograms) with J; obviously the transition width increases with J.   }
\end{figure}  
Figure 2 shows the representative raw DTA thermograms recorded under different bias current. With the increase in current, the peak broadens while the area under the peak decreases. Since the Joule heating near the transition zone is negligible, with the increase in J, the baseline of the DTA trace does not change at all.\cite{note} Therefore, no compensation was necessary for detecting the peak and its variation under enhanced current density. The latent heat has been calculated by subtracting the background by an appropriate technique and identifying the peak area properly. The errors in such estimation are calculated to be $\pm$0.5$\%$ in the case of transition temperatures and $\pm$10$\%$ in the case of latent heat. In Fig. 3(a) we show the variation in the latent heat ($\textit{L}$) of transition with $\textit{J}$ while in Fig. 3(b) the variation of the transition temperatures is shown. There is a slight history effect as the transition temperatures and peak area differ a bit between heating and cooling cycles. This is not because of any intrinsic effect (e.g., due to slower phase transition dynamics or metastability) as discussed later, but could be due to slight impurity in the inert atmosphere (flowing nitrogen) maintained during the experiment. In fact, application of different heating/cooling rates did not result in any significant shift in the transition temperatures. The $\textit{L}$ is found to decrease gradually with the increase in $\textit{J}$ and reach zero at $J_C$ $\sim$50 $A/cm^2$ [Fig. 3(a)]. This pattern of variation of $\textit{L}$ with $\textit{J}$ is consistent with the variation of the extent of hysteresis ($\Delta T$) in the  $\rho-T$ plot around T$_{OO}$ between the heating and cooling runs; $\Delta T$ too decreases with the increase in $\textit{J}$ and vanishes at $J_C$ [Fig. 3(a), inset]. The transition zone (marked by the onset, peak, and end temperatures) observed in Fig. 3(b) also broadens with $\textit{J}$. It is important to mention here that the orbital order-disorder transition even in a very high quality single crystal of LaMnO$_3$ under nearly zero $\textit{J}$ is actually a broadened first order transition.\cite{Zhou} No thermodynamic evidence for strictly first order transition has so far been reported. In comparison, compelling thermodynamic evidence for the first order transition has been gathered using local magnetization measurement in the case of vortex lattice melting in high-T$_C$ superconductor. A step-like rise in magnetization could be noticed within a temperature range of $\sim 3$ $mK$ around the vortex lattice melting line.\cite{Zeldov} It has also been shown that disorder can broaden the first order transition.\cite{Soibel} In the present case, of course, the broadening of the transition even under very low current density could be because of intrinsic inhomogeneity or disorder due to the presence of orbital domains. Therefore, we compare the latent heat as a function of $\textit{J}$ only in the sense of noting the relative variation. The nature of the transition actually broadens progressively and finally becomes broader than the resolution of the instrument. At that point, the isolation of the peak area from the baseline is no longer possible and the estimated $\textit{L}$ reaches zero. Using a calorimeter of higher sensitivity or a local calorimeter one could possibly resolve the peak. 
\begin{figure}[!h]
\centering
\includegraphics[scale=.35]{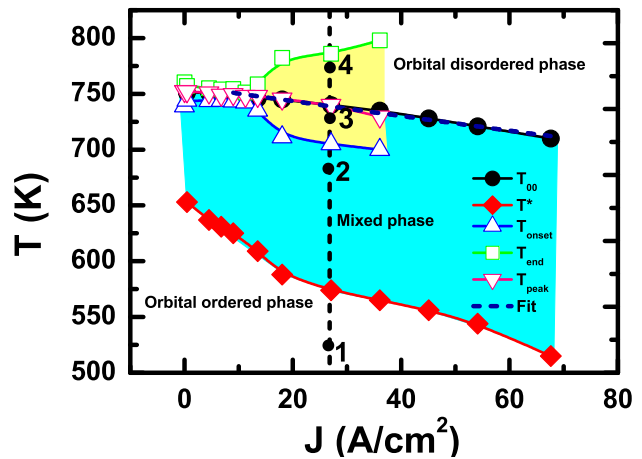}
\caption {(color online) The phase diagram of electric current driven orbital order-disorder transition on the $J$-$T$ plane: the transition width increases with the increase in $J$. The peak width of the DTA data has been superimposed on the phase diagram. It shows that the DTA peak width is quite smaller in comparison to the transition width identified from the electrical resistivity data. For clarity, the error bars have been omitted here. The points at which the data of relaxation of resistivity have been presented in Fig.5 are shown as 1,2,3, and 4.}
\end{figure}

Using the data of transition temperatures as a function of bias current, obtained both from electrical and calorimetric measurements, we construct a phase diagram on the $J-T$ plane (Fig. 4). The phase diagram indicates three regions: (i) ordered, (ii) disordered, and (iii) transition. It is clear from the figure that the transition zone broadens progressively with current density. Interestingly, the peak broadening, observed in DTA data under enhanced $\textit{J}$, covers a rather narrow portion of the $T^*-T_{OO}$ transition zone identified from the electrical resistivity data. It shows that though the electrical measurement senses the onset of transition at $T^*$, the calorimetric measurement senses the onset at a much higher temperature closer to the thermodynamic $T_{OO}$. Earlier work\cite{Maris} on evolution of crystallographic structure across $T^*-T_{OO}$, on the contrary, reveals that the anomalous structural distortion sets in at $\sim T^*$ itself. The reason behind discrepancy between the onset points identified from the crystallographic and resistivity data and those recorded from the calorimetric data could be the difference in sensitivity of the probes. The calorimeter sensitivity is $\le$1 $\mu$W whereas the sensitivity of the nanovoltmeter used for recording the voltage drop and hence resistance of the sample is $\le$10 nV. Therefore, while the electrical resistivity and x-ray diffraction experiments could sense the nucleation of the orbital disordered phase and hence record accurately the onset of transition, the calorimeter could record the onset of transition only when the orbital disordered phase has grown beyond a critical size and hence at a higher temperature.  
\begin{figure}[!htp]
\centering
\includegraphics[scale=.43]{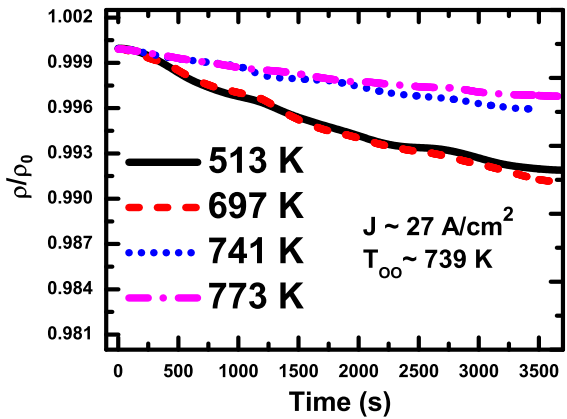}
\caption {(color online) A representative plot of time dependence of normalized electrical resistance at different temperatures under a bias current density $\sim$27 $A/cm^2$. The temperatures and the bias current are chosen in such a way so that different regions of the phase diagram - ordered, disordered, and transition - can be accessed. The extent of relaxation appears to be negligible in all the cases indicating stability of the phases. On the contrary, 2-4\% decay in resistance could be noticed\cite{Mondal} in the case of metastable orbital ordered phase in nanoscale ($\leq 20$ nm) LaMnO$_3$. } 
\end{figure}

In order to probe further the characteristics of these three regions of the $J-T$ phase diagram, especially, whether or not the slight history effect observed in DTA data around the transition is due to intrinsic metastability of the phases at the transition zone, we have measured the relaxation of resistivity. For crossing the boundaries of the transition zone along a constant $\textit{J}$ line, we selected four points (shown in Fig.4) with different temperatures. The relaxation measurements were carried out by raising the temperature of the crystal to the desired point and then applying the requisite current density. We reached the points separately by raising the temperature and field from room temperature and zero-field. The resistivity data were recorded over a time span of $\sim$50ms to 3600s at an interval of $\sim$50ms after stabilizing the sample temperature at a given point within less than 0.01 K. We repeated such measurements along other constant $\textit{J}$ lines too. In Fig. 5, we show the relaxation characteristics observed at points 1, 2, 3, and 4. The characteristics are representative of all the points at which similar relaxation measurements were carried out. We observe virtually no time dependence of resistivity corresponding to the points 1, 2, 3, and 4 signifying stability of not only orbital ordered and disordered phases but also of the mixed phase within the transition zone. Therefore, the states at transition zone are not metastable. The slight difference observed in the transition temperatures between heating and cooling cycles (Fig. 3) cannot be due to metastability and consequent emergence of transient states around the transition. The transition is thermodynamic and the transition kinetics is not accessible within the laboratory time window $\sim 10^{-3}-10^4$s at all the temperatures. Probing of evolution of crystallographic structure\cite{Maris} within and around the transition zone shows that O'  orthorhombic structure ($c/\sqrt{2}< a < b$; a, b, c are lattice parameters) of the orbital ordered phase evolves into a mixed O'  and O orthorhombic phase within the transition zone and finally into a pure O orthorhombic phase ($c/\sqrt{2}\sim a \sim b$) above the transition zone. By taking into consideration the evolution of the structure within the transition zone along with the relaxation data of resistivity, it is possible to conclude that the mixed O'+O phase does not have any temporal fluctuation within the laboratory time window. On the contrary, orbital ordered phase in nanoscale ($<$20 nm) LaMnO$_3$ is found to be metastable with 2-4\% decay in resistance with time and irreversible order-disorder transition within similar time scale.\cite{Mondal} Emergence of transient metastable states with lifetime ($\sim 10^{-3}-10^4$s) comparable to the laboratory time scale has been observed by Kalisky \textit{et al}.\cite{Kalisky} around the vortex solid-solid phase transition as well, in a high-T$_C$ superconductor.  

The decrease in T$_{OO}$ as well as latent heat and hysteresis associated with the transition possibly result from electric current driven depinning of orbital domains. The orbital ordered phase even in stoichiometric LaMnO$_3$ is not a continuum under zero electric current. It contains domains due to interaction with intrinsic electronic/lattice defects, strain etc.\cite{Ahn,Nelson} Like in the case of charge density waves in solids,\cite{Gruner} these intrinsic defects act as pinning centers for orbital domains as well. In fact, orbital order-disorder transition takes place via depinning of orbital domains.\cite{Uchida} In the orbital disordered state, the short-range order, with high mobility and hence temporal fluctuations, prevails.\cite{Qiu} Application of enhanced current density (bias field) leads to electromigration of defects\cite{Sorbello} which, in turn, can give rise to field driven depinning transition. The depinning transition of charge density waves as well as vortex lattice in high-T$_C$ superconductors has been thoroughly studied.\cite{Vinokur} It has been shown that depending on the concentration of pinning centers and applied force, the depinning of the charge density waves can either follow a two-stage or a single-stage process. If the concentration of defects is strong, the domains start sliding plastically under a small force which finally gives way to a sharp transition into a coherent collective movement in 3D under higher force. On the other hand, in the case of weak disorder or low concentration of defects, the depinning transition becomes a continuous process and yields a coherently moving collective state continuously. The weak disorder model predicts that the depinning transition yields an exponential variation of the transition energy scale (e.g., transition temperature) with the applied force.\cite{Monceau} Interestingly, this model is found to be valid in the present case. It has been observed that the T$_{OO}$ versus current density J pattern (Fig. 4) follows closely the model T$_{OO}$(J)/T$_{OO}$(0) = exp[-J/J$_0$] (dashed line, Fig. 4), except at a very low J (J$_0$ is a constant here and T$_{OO}$ is the transition temperature under zero field). Validity of this model in the present case provides an indirect support for the conjecture of current density (bias field) driven depinning of orbital domains. Because of variation in the depth of the potential well of defects, the screening of direct force for electromigration will vary\cite{Lodder} which, in turn, is expected to give rise to inhomogeneous depinning. This inhomogeneous depinning is possibly the origin of broadening of the orbital order-disorder transition zone and drop in the latent heat. 

Can there be any other origin for the influence of electric field on orbital order in LaMnO$_3$? The d-type orbital order in LaMnO$_3$ cannot give rise to a electric dipole moment. It only produces a higher order quadrupole moment\cite{Sartbaeva} which cannot couple linearly with the applied electric field. Even the domain boundaries of orbital order cannot be intrinsically charged. Therefore, unlike charge order, microscopically, long-range orbital order in LaMnO$_3$ should not be influenced by electric field. Whether unleashing of charge carriers via $Mn^{3+} \rightarrow  Mn^{2+} + Mn^{4+}$ disproportionation reaction under field could then be the origin of such an effect? The generation of mobile charge carriers under field would have given rise to even more dramatic effect. Therefore, it seems, apparently, that these effects are not really playing any significant role here. Study of orbital domain structure under external electric field using spatio-temporally resolved resonant x-ray scattering data can offer direct proof of the orbital domain depinning transition under field. This is beyond the scope of this paper.

\section{Summary}
In summary, we observe significant influence of electric current density (bias field) on orbital order-disorder transition in a single crystal of pure LaMnO$_3$. The transition temperature, hysteresis, and the latent heat of transition decrease monotonically with the increase in the current density; the transition width, on the contrary, increases. Finally, at a critical current density, the latent heat becomes zero. The states at the ordered, disordered, and transition zones are all found to be thermodynamically stable within the laboratory time scale. This current density driven orbital order-disorder transition possibly originates from the field driven depinning of orbital domains via electromigration of intrinsic defects. 

We acknowledge helpful discussion with R. Nath, A.K. Raychaudhuri, T. Saha-Dasgupta, and D.I. Khomskii. One of the authors (PM, first author) acknowledges financial support from CSIR during this work.

\end{document}